\documentclass[3p,times,twocolumn]{elsarticle}

\usepackage{ecrc}

\volume{00}

\firstpage{1}

\journalname{Nuclear Physics B Proceedings Supplement}

\runauth{E. Mocchiutti}

\jid{nuphbp}

\jnltitlelogo{Nuclear Physics B Proceedings Supplement}

\usepackage{amssymb}

\usepackage[figuresright]{rotating}

\begin{document}

\begin{frontmatter}



\dochead{}

\title{Direct detection of cosmic rays: through a new era \\of precision measurements of particle fluxes.}


\author{E. Mocchiutti}
\ead{Emiliano.Mocchiutti@ts.infn.it}

\address{INFN, Sezione di Trieste, I-34149 Trieste, Italy}

\begin{abstract}
In the last years the direct measurement of cosmic rays received a push forward by the possibility of conducting experiments on board long duration balloon flights, satellites and on the International Space Station. The increase in the collected statistics and the technical improvements in the construction of the detectors permit the fluxes measurement to be performed at higher energies with a reduced discrepancy among different experiments respect to the past. However, high statistical precision is not always associated to the needed precision in the estimation of systematics; features in the particle spectra can be erroneously introduced or hidden. A review and a comparison of the latest experimental results on direct cosmic rays measurements will be presented with particular emphasis on their similarities and discrepancies.
\end{abstract}


\end{frontmatter}


\section{Cosmic rays direct detection}
Cosmic rays were discovered about one century ago by Victor Hess. Hess was awarded with the Nobel prize in 1936 for his studies, but he was never able to actually perform a direct detection of the cosmic rays due to the technological limitations of balloon flights and of the detectors at his times. Indeed he was able to measure the amount of secondary produced radiation in the atmosphere, that is to study the development of cosmic ray showers generated by primary cosmic rays. It took several years to understand that the main component of cosmic rays is made of protons with a steeply falling flux as function of energy. The study of cosmic radiation and its interaction with the Earth atmosphere led to the discovery of new particles and set the basis for the experimental particle physics that is carried out today at accelerators. 

The origin, acceleration and propagation mechanisms of charged particles traveling in the Space have been the main topics in the studies of cosmic radiation since its discovery. In the aim of solving these puzzling issues, in the 80s and 90s a massive campaign of experiments was carried out on stratospheric balloon flights and small satellites. With increasing knowledge on the cosmic rays, it began to be clear that it is very difficult to provide a satisfactory and self-consistent global model. Sources types, their chemical composition and their inner dynamics, acceleration processes, and propagation through the interstellar matter and in the heliosphere affect the shape and the composition of the fluxes measured at Earth. 

\begin{figure}[!htb]
\begin{center}
\includegraphics[width=7.5cm]{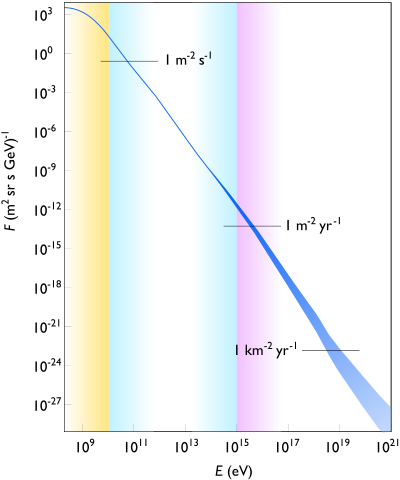}
\caption{All-particle cosmic ray flux.}
\label{fig1}
\end{center}
\end{figure}

The cosmic ray all-particle spectrum is shown in figure~\ref{fig1} \cite{lafebre}. 
Most experiments agree, at least qualitatively, that the spectrum consists of at least three regions. At the lowest energies, from tens of MeV to tens of GeV (yellow band in figure), the particles coming from the interstellar space are deflected and influenced by the magnetic region generated by the Sun, the heliosphere. As a consequence, the observed spectrum is flattening. At higher energies, instead, the direct measurements of cosmic rays represent the interstellar flux and composition. For energies between tens of GeV and about $10^{15}$~eV, the spectrum can be fit with a power law with slope $\sim$2.7. Due to such a steep spectrum, with current technology, a direct measurement (cyan band in figure) is possible only up to about $10^{15}$~eV, the so-called ``knee''. Beyond the knee (purple band), the slope grows to $\sim$3.1. Only indirect measurements are possible by exploiting the atmosphere as a large calorimeter and by making use of ground based detectors. At the highest energies particles have energies comparable to the Greisen-Zatsepin-Kuzmin limit (GZK cutoff), which occurs at about 5$\times$10$^{19}$~eV.

At the end of the 90s, experimental cosmic ray direct measurements were limited to few hundreds of GeV for protons and helium nuclei (major component of cosmic radiation) and to few tens of GeV for antiparticles. Due to the limited statistics and to quite large systematic uncertainties it was still not possible to answer to many fundamental questions concerning the cosmic rays. As a consequence, the experimental study of cosmic rays took three paths that are still effective. The first research line aims to push the direct measurements at the highest energies, possibly reaching the knee, in order to study sources and acceleration mechanisms. The second research line is dedicated to study the chemical composition of cosmic rays, measuring highly charged nuclei spectra, with the aim of understanding the source material, dust and gas, the nucleosynthesis and the propagation of cosmic rays in the interstellar medium. The third path, finally, is dedicated to the study of the rare antiparticle and anti-matter component, trying to search for signal of the elusive dark-matter, set anti-matter limits and understand the matter-antimatter asymmetry in the Universe.

Depending on the research line, different platforms and detection techniques have been adopted. In the following, I will describe the latest missions conducted on stratospheric balloons, satellites and on the International Space Station (ISS) while discussing the main physics results obtained in the recent years. I will categorize the results by type of particles and their role in the cosmic ray ``standard model''.
\begin{figure}[htb]
\begin{center}
\includegraphics[width=7.5cm]{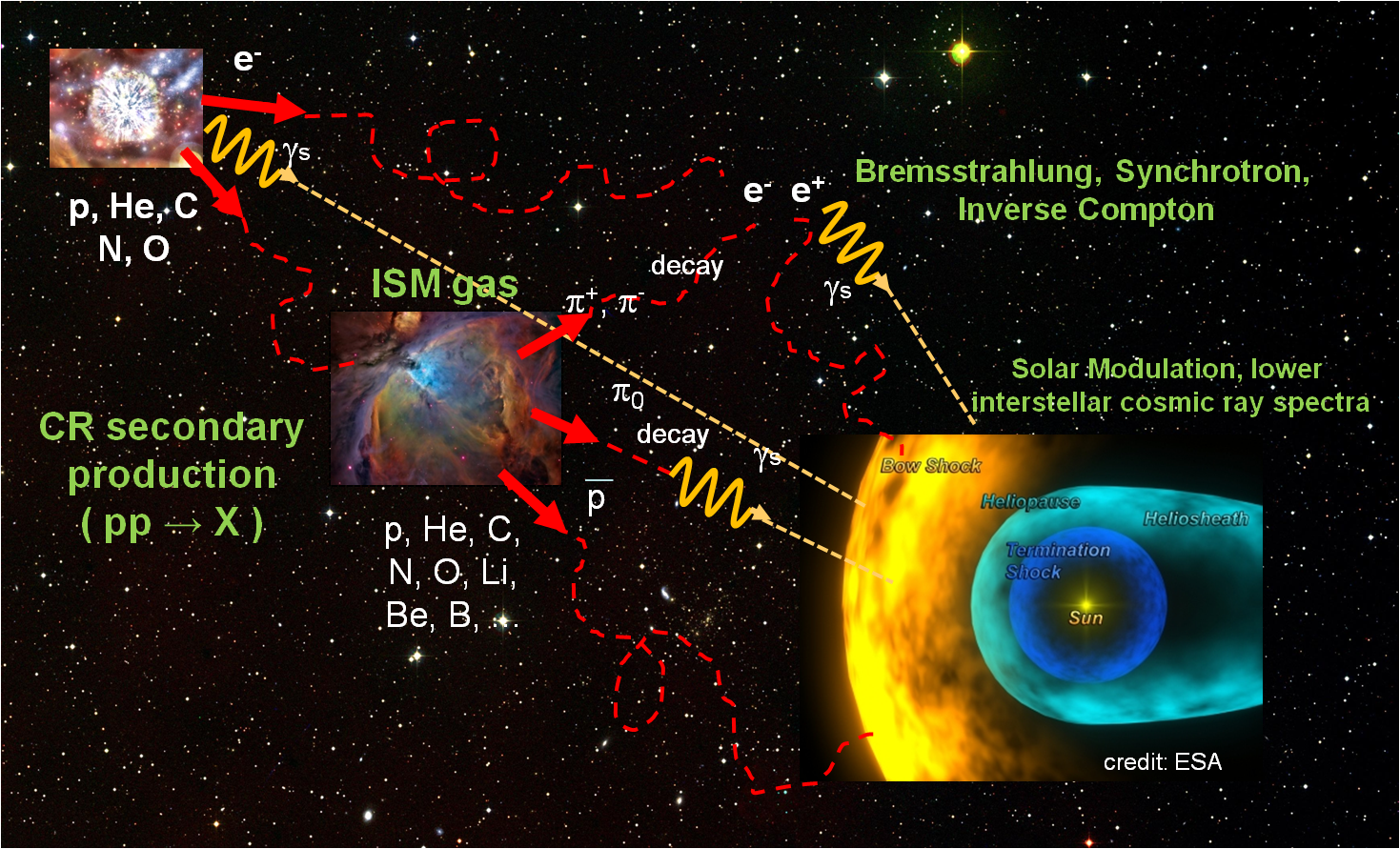}
\caption{Schematic representation of the cosmic ray ``standard model''.}
\label{fig2}
\end{center}
\end{figure}
With ``standard model'', figure \ref{fig2}, I refer to the idea that protons, helium nuclei, electrons and highly charged stable nuclei are accelerated as primary cosmic rays by supernoavae explosions. Deflected by the galactic magnetic field, these particles reach the solar system where they must enter the heliosphere and the Earth magnetic field (magnetosphere) to be detected by our detectors. During their travel, primary cosmic rays can interact with the interstellar matter gas and they can generate any kind of particle, secondary cosmic rays. In this model, fragile nuclei and antiparticles observed in the cosmic rays are only of secondary origin. Source distribution is supposed to be isotropic in space and time, and particles are assumed to gain energy via the second order Fermi particle acceleration process. The resulting fluxes in this model are smooth and steady.

\section{Protons and helium nuclei spectra}
\label{pandhe}
Protons and helium nuclei are the most abundant cosmic ray component and hence, from the experimental point of view, the particle selection is significantly easier with respect to the one needed for studying the rare component.

Measurements of primary cosmic-ray proton and helium nuclei spectra have been performed over the years using different techniques: magnetic spectrometers and RICH detectors have been used for energies up to 1 TeV/n, while calorimetric measurements extended to higher energies. The majority of these results, especially concerning the high-energy ($\simeq 1$ GeV) part of the spectra, were obtained by balloon-borne experiments. Recently, however, two space experiments presented their proton and helium nuclei results: PAMELA and AMS-02.

\begin{figure}[!htb]
\begin{center}
\includegraphics[width=7.5cm]{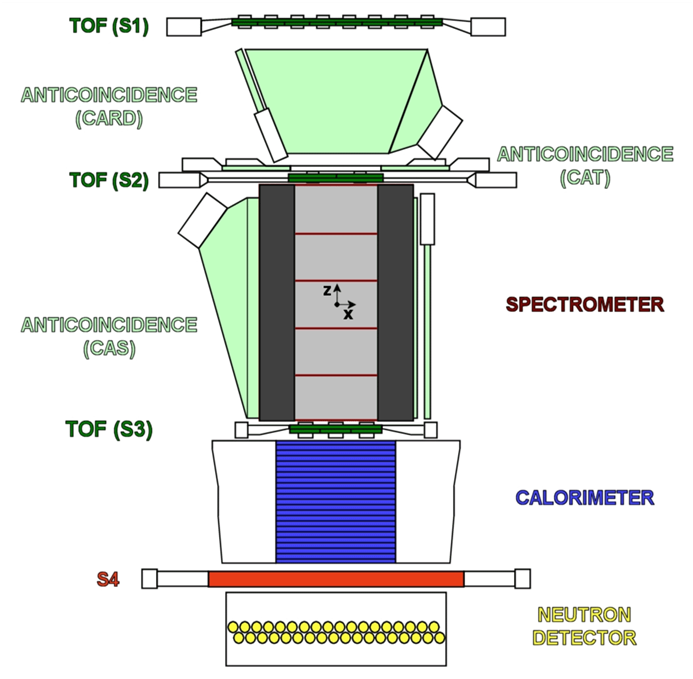}
\caption{Schematic view of the PAMELA apparatus.}
\label{fig3}
\end{center}
\end{figure}

PAMELA (Payload for Antimatter Matter Exploration and Light-nuclei Astrophysics) is a satellite-borne experiment designed to study charged particles in the cosmic radiation. PAMELA was launched into space by a Soyuz-U rocket on June 15$^{th}$, 2006 from the Baikonur cosmodrome (Kazakhstan) ~\cite{pic07}.  The primary scientific goal of the mission is the study of the antimatter component of the cosmic radiation up to energies of hundreds of GeV. A schematic view of the apparatus is shown in figure~\ref{fig3}. 
The central part of the apparatus is a magnetic spectrometer, consisting of a permanent magnet and a silicon tracker. The main task of the magnetic spectrometer is to measure particle rigidity $\rho = pc/Ze$ ($p$ and $Ze$ being respectively the particle momentum and charge, and c the speed of light) and ionization energy losses ($dE/dx$). An electromagnetic calorimeter is installed below the magnetic spectrometer and it is used for the particle identification~\cite{boe06}. The Time-of-Flight (ToF) system of the experiment provides a fast signal for triggering the data acquisition in the PAMELA subdetectors and performs up to 12 independent measurements of the particle velocity, $\beta = v/c$. By measuring the particle velocity the ToF system discriminates between down-going and up-going splash albedo particles, thus enabling the spectrometer to establish the sign of the particle charge. The ToF system also provides 6 independent $dE/dx$ measurements, one for each scintillator plane. The aim of the anticoincidence (AC) system is to identify, in the offline analysis, false triggers and multi-particle events, generated by secondary particles produced in the apparatus. Below the electromagnetic calorimeter, a single square plastic scintillator (S4) acts as a shower-tail catcher and is used to generate a high energy trigger signal for the underlying neutron detector (ND). The purpose of the neutron detector is to complement the electron-proton discrimination capabilities of the calorimeter by detecting the increased neutron production associated with hadronic showers in the calorimeter compared with electromagnetic ones.

The Alpha Magnetic Spectrometer (AMS-02)~\cite{agu13} is a particle physics experiment module that is mounted on the International Space Station. It is designed to measure antimatter in cosmic rays and search for evidence of dark matter. Conceptually similar to PAMELA, AMS has a much larger acceptance (about twenty times with respect to PAMELA) and redundant detectors for particle identification. 
\begin{figure}[!htb]
\begin{center}
\includegraphics[width=7.5cm]{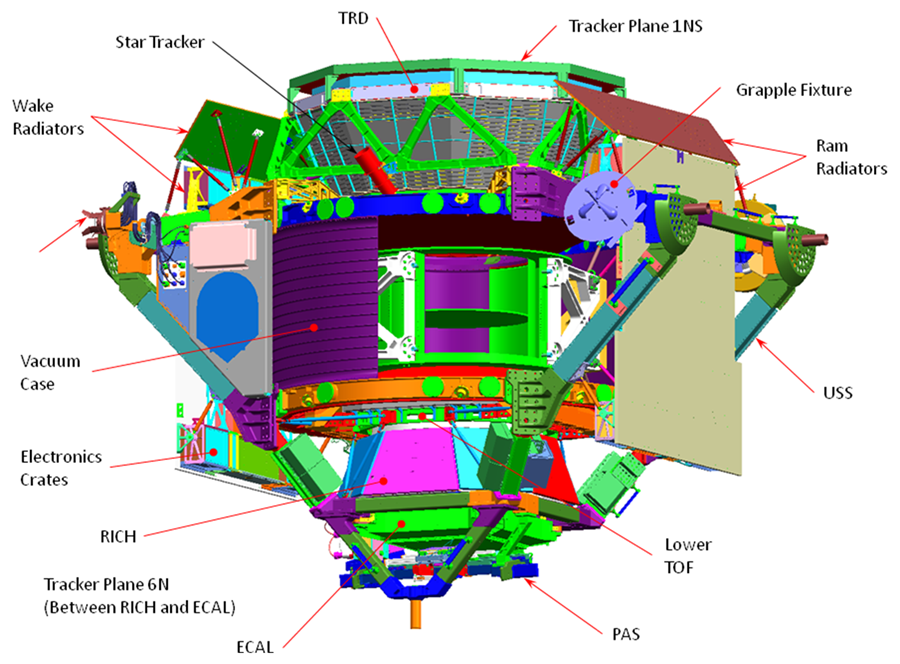}
\caption{Schematic view of the AMS-02 apparatus.}
\label{fig4}
\end{center}
\end{figure}
For instance, figure \ref{fig4}, a transition radiation detector (TRD) installed above the permanent magnet permits to identify electrons and positrons among the antiprotons and protons and a ring-imaging Cherenkov (RICH) detector measures with high precision the cosmic ray velocity. Time of flight (ToF), anti-coincidence system (ACC) and silicon tracker (Tracker) have the same functionalities as in the PAMELA apparatus.

The proton and helium nuclei spectra were published by the PAMELA collaboration~\cite{adr11a} and were presented, as a preliminary work, at the International Cosmic Ray Conference (ICRC) in 2013 by the AMS collaboration~\cite{ams13a,ams13b}. Since the latter are only preliminary results, only a preliminary and not conclusive comparison between the two set of data can be carried out\footnote{All AMS-02 preliminary results presented in this paper were extracted, using plot digitizer open source software, from freely available figures at the AMS-02 website (http://www.ams02.org) downloaded in September 2013.}. 
\begin{figure}[!htb]
\begin{center}
\includegraphics[width=7.5cm]{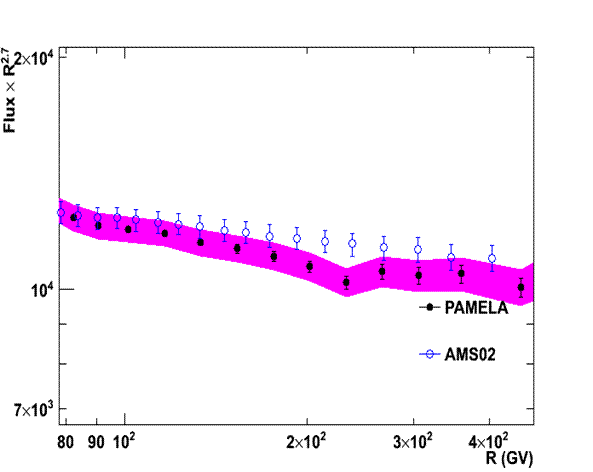}
\caption{Comparison between proton fluxes as function of the rigidity measured by PAMELA and by AMS-02.}
\label{fig5}
\end{center}
\end{figure}
Figure~\ref{fig5} shows the proton fluxes measured by PAMELA and AMS-02 in the rigidity range between about 80 and 450 GV. PAMELA error bars represent statistical errors, while the band represents the statistical uncertainties described in~\cite{adr11a} quadratically summed. AMS-02 error bars represent both the statistical and systematic uncertainties quadratically summed. 

As a general comment, these results represent a step forward with respect to the ones obtained by the experiments of the previous generation:
\begin{itemize}
\item the measured fluxes extend over a wide energy range, from the hundreds of MeV to the TeV region;
\item the collected statistics is huge, order of millions of events;
\item the precision in the energy measurement combined with the high statistics permits to adopt a very thin binning in energy;
\item the systematic uncertainties are contained at the order of the percent level, to be compared with the about 10\% systematic uncertainties of the previous generation of experiments.
\end{itemize}
However it is evident from figure~\ref{fig5} that the systematic uncertainties are dominating the measurements, being much greater than the statistical errors. This issue nullifies the benefits of having high statistics and a very precise energy measurement, at least in most of the energy window under investigation.

\subsection{Systematic uncertainties: sources and estimation}
Since systematics play such an important role in this type of measurements, I will dedicate this section to discuss their origin and meaning, focusing on space spectrometer detectors. 

Systematic uncertainties represent the level of knowledge of the detector: they depict what we know we don't know about our apparatus. Being so, the systematic uncertainties cannot be ``measured'', only estimated, always remembering that the perfect knowledge of the experimental setup is not achievable and a small, hopefully, level of unknown remains. In the case of the flux estimation performed with PAMELA and AMS-like detectors, a not-comprehensive list of systematic uncertainty sources includes:
\begin{itemize}
\item estimation of selection efficiencies;
\item estimation of residual sample contamination (if any);
\item measurement of the particle energy.
\end{itemize}
Particle selection is the combination of several cuts on distribution of different observables measured with different detectors. Usually a first cleaning of the full data set is required in order to reject random coincidences and spurious events, and to select high quality reconstructed events. Then dedicated selections are applied in order to distinguish the particle type. Hence, even in the case of the most abundant particles in the cosmic radiation, the initial sample undergoes a set of cuts which efficiency must be estimated in order to measure the particle spectrum. The efficiencies estimation is, usually, performed independently for each selection and in cascade in order to avoid correlation problems. Efficiencies can be determined making use of experimental data sample selected with independent detectors or by means of simulations. Flight data sample can potentially bring uncertainties due to energy, acceptance or time dependences and they can be possibly be contaminated by other type of particles. Moreover, experimental data sample usually can cover only a portion of the energy window under investigation and extrapolations to the whole range are needed. Simulation data samples are easier to handle and they are fully under control for what concern the type of particles and detector configuration. However, the simulation cannot include what it is not know about the detector, or what is unknown in the description of physics processes in our apparatus. Eventually, the determination of selection efficiency consists in a combination and a comparison of different methods, samples and detectors. The spread of the results around the value that is used for the flux determination is used as an estimation of the selection efficiencies systematic uncertainties. 
\begin{figure}[!htb]
\begin{center}
\includegraphics[width=7.5cm]{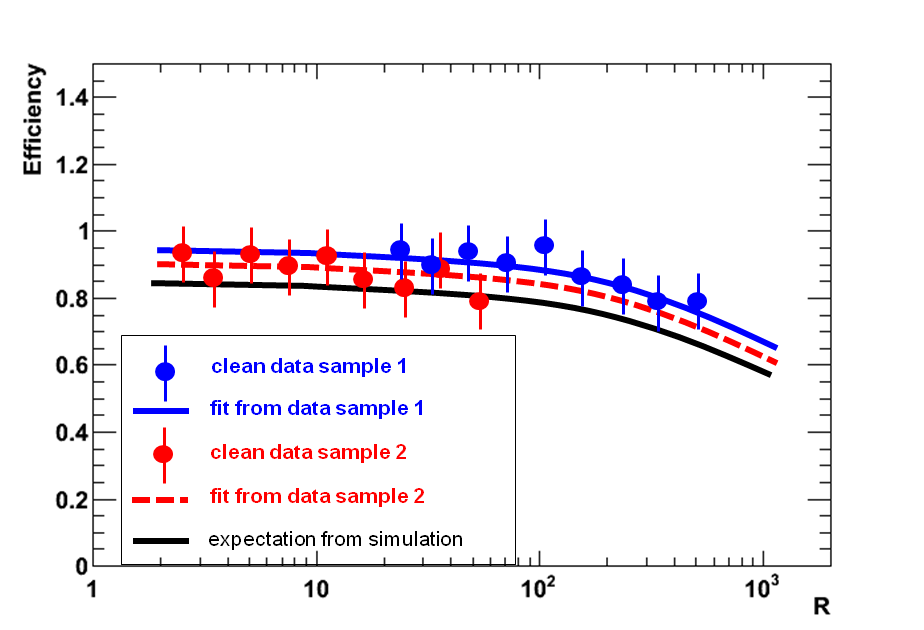}
\caption{Example (fictious) of the differences between three methods of efficiency estimation.}
\label{fig6}
\end{center}
\end{figure}
Figure~\ref{fig6} shows a fictious example of differences in the determination of a selection efficiency. In the figure the results coming from two possible data sample are plotted. They are supposed to be obtained from experimental data which cover different energy regions (red and blue dots). The output expected using the simulation (black line) is shown together with the fits of these distributions (red and blue lines), obtained assuming a given shape of the data. Assuming that there is no reason to doubt of any of the results and since there is a small but not negligible discrepancy between the three estimations, eventually one of the curves is chosen for the flux measurement and the relative difference with the others is an estimation of the systematic uncertainty for this efficiency. It is difficult to determine the effect of this type of systematic uncertainty on the resulting flux. The quoted systematic error bar or band can mean a systematic solid shift of the flux but it can also hide distortions or enhanced fluctuations around the true value.

The same reasoning applies to the estimation of residual sample contamination. In the case of high identification power of the apparatus, the contamination is rendered negligible by making use of proper selections. If a spectral analysis is used, the background must be estimated and this estimation can be done, as in the case of selection efficiencies, by means of real data samples or simulation. 

\begin{figure}[!htb]
\begin{center}
\includegraphics[width=8.5cm]{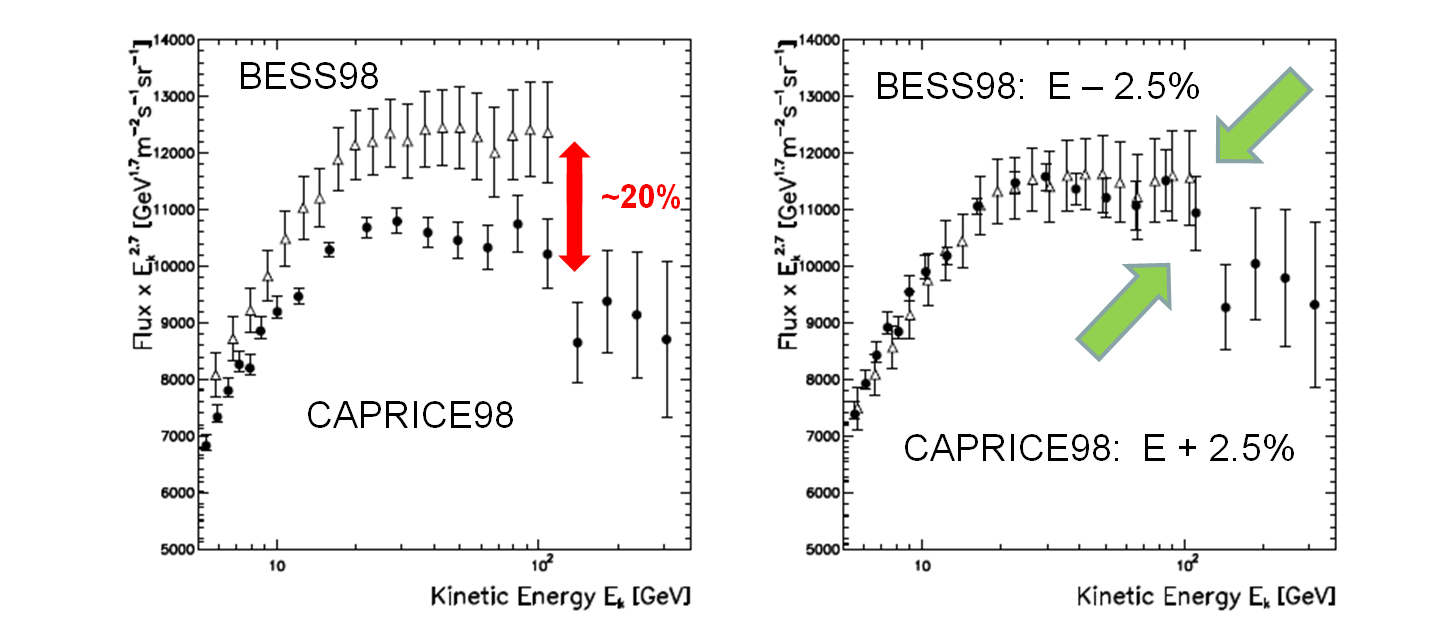}
\caption{Effects of energy measurement uncertainty on the CAPRICE98~\cite{boe03} and BESS98~\cite{hai04} proton fluxes~\cite{moc03}.}
\label{fig7}
\end{center}
\end{figure}

The measurement of the particle energy in case of magnetic spectrometer is also a delicate issue. The actual measurement is the position measurement on several detector planes placed in a magnetic field. From the position measurements the deflection of the particle trajectory is determined. The rigidity, momentum over charge, is hence given by the inverse of the deflection. Kinetic energy is finally obtained by converting the rigidity of the particles once the mass and the value of the charge is known. The effects of systematic errors in the energy measurements are, again, difficult to determine. Usually a distortion of the data around the true value has to be expected. For example, in figure~\ref{fig7}
 on the right, the effects on the CAPRICE98~\cite{boe03} and BESS98~\cite{hai04} proton fluxes of a total shift in energy of about 5\% are shown~\cite{moc03}. A difference of about 20\%, left panel, is compensated by a relatively small change in the energy scale. Notice that this compensation is mostly due the representation of fluxes multiplied by the energy to a certain power. Effects due to deflection systematic errors are more subtle, bringing an intrinsic distortion as function of the energy.

Summarizing, it is difficult to estimate the effects of systematic uncertainties on the measured fluxes and the systematic error bars or bands represent the best estimation of our understanding of the detectors. These uncertainties can potentially hide or create features and distortions of the measured spectra. In addition they can result as a global normalization problem between different data sets (i.e. a solid shift of the spectrum). It is therefore of extreme importance to remind the main differences between systematic and statistical uncertainties:
\begin{itemize}
\item statistical uncertainties are measured, they depends on the number of collected events, while systematic uncertainties can only be estimated;
\item statistical uncertainties imply a scattering of the data around the true value, systematics can have a different effect depending on their type and estimation;
\item the meaning of statistical uncertainties is the same for any experiment, while systematics are experiment-dependent and even the same type of systematic could have different effect on the fluxes depending on different detectors;
\item statistical uncertainties are time independent, systematics represent the knowledge of the detector which behavior can vary with time (e.g. due to aging or exposure to radiation);
\item statistical analysis must be used for statistical error propagation; in the case of systematic uncertainties statistical analysis could introduce bias in their estimation, in fact correlation between different systematics is difficult to settle since they represent unknown effects;
\item statistical errors can be reduced by collecting more data, while systematic uncertainties are difficult to lower. Even if the knowledge of the detector can improve by further analyzing and collecting data it is also true that an experimental apparatus changes as function of time hence limiting our capabilities in the control of systematic effects.
\end{itemize} 
Considering both statistical and systematic uncertainties, the PAMELA and AMS-02 proton and helium nuclei results are in an overall good agreement. It must be noticed also that the differences in the fluxes measurement between the two experiments is at the percent level, while the agreement of previous generation detectors was much worse, at the level of 15, 20\%. 

\subsection{Proton to helium ratio}
As discussed previously, the latest proton and helium nuclei results from PAMELA and AMS-02 are dominated by systematic uncertainties from the GeV region up to hundreds of GeV. While disappointing, this is a natural consequence of the main experimental goal of the two experiments: the study and detection of the anti-particles, anti-matter and dark matter in the cosmic rays. A huge cosmic ray statistics is needed to detect such an elusive signal. Moreover, to perform indirect dark matter searches it is necessary to push the energy measurement to the highest energies, above the TeV region. By setting these objectives, a natural by-product is the measurement of the most common proton and helium nuclei but at the price of having a huge statistics and an experiment that is not tuned to be ``free'' of systematic uncertainties in the ``low'' energy window. 

Interestingly, the measurement of flux ratios is much more solid from the point of view of systematic uncertainties, since in that case most of the selections are the same and the associate systematics cancel out. It is therefore not surprising that the proton over helium ratio 
\begin{figure}[!htb]
\begin{center}
\includegraphics[width=7.5cm]{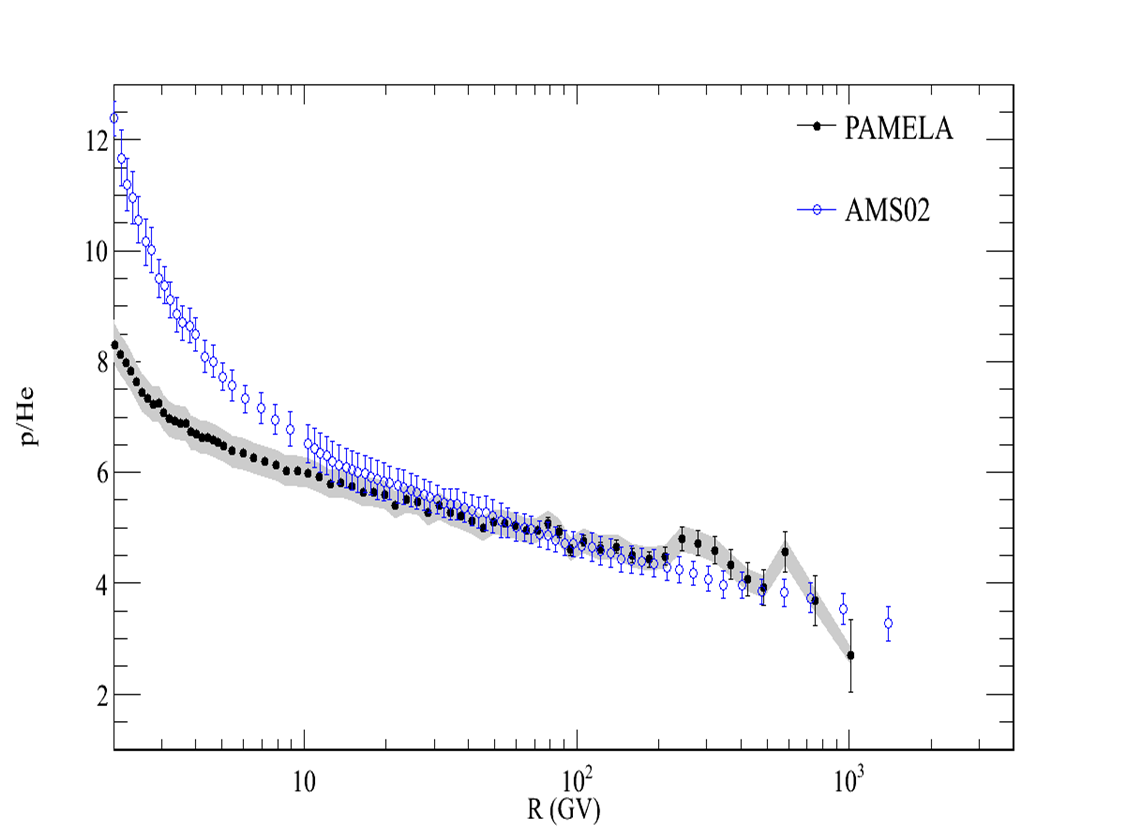}
\caption{Proton over helium nuclei ratio as function of the rigidity, comparison between PAMELA published results and preliminary AMS-02 data. Band and error bars meaning are the same of figure~\ref{fig5}.}
\label{fig8}
\end{center}
\end{figure}
as measured by PAMELA and AMS-02, figure~\ref{fig8}, shows an excellent agreement. Below some tens of GeV there is an expected difference due to the solar modulation at the time of the acquisition. At high energy both the measurements show that proton and helium nuclei fluxes have a different spectral index. This result challenges the standard model paradigm with important implications on the origin, acceleration and possibly the propagation of these two types of particles.

The spectral features observed by PAMELA in the fluxes are not present in the proton to helium ratio. More comments about the fluxes shape will be made in section~\ref{chemic}, after the introduction, in the following sections, of the results from other experiments.

\section{Low energy proton measurements}
Some of the limitations of balloon-borne missions carried out in the 90s have been overcome over the last years by means of Long or Ultra-Long Duration Balloon Flights (LDB or ULDB) flown out of locations in Alaska and Antarctica. LDB and ULDB balloons are made out of special materials and can stay aloft up to 100 days.

The main goals of balloon experiments realized in the last decade are: the study of low energy antimatter in cosmic rays using magnetic spectrometers; the measurement of the cosmic-ray composition up to the highest energies using large acceptance detectors with excellent charge identification capabilities and good energy measurement resolution obtained using calorimetric or velocity measurements.

Polar flights from the BESS collaboration in 2004 and 2007 were very successful and a large data set was collected. The aims of the experiment are precise measurements of the low-energy antiproton spectrum and search for antimatter in the cosmic rays.
\begin{figure}[!htb]
\begin{center}
\includegraphics[width=7.5cm]{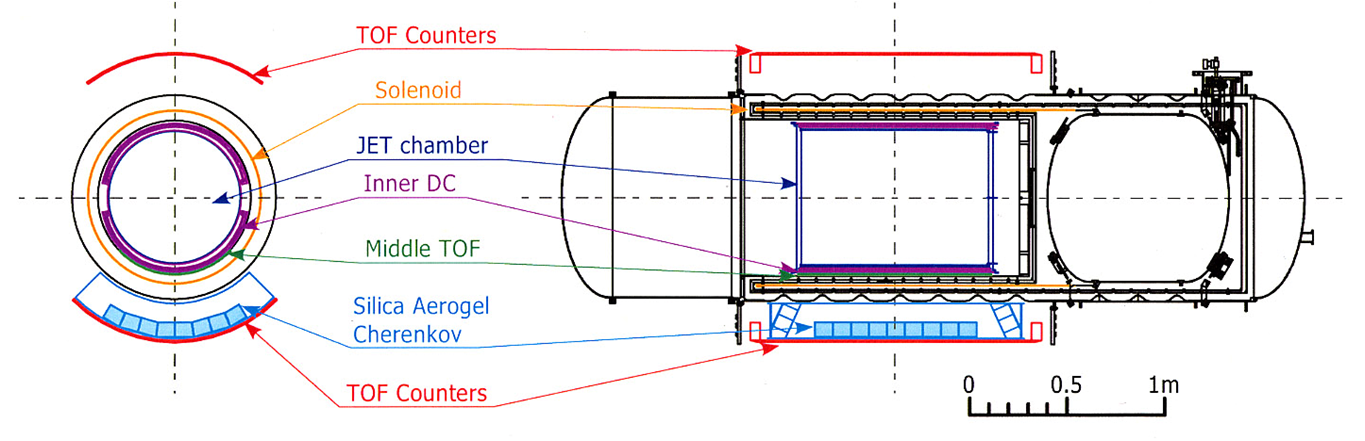}
\caption{Schematic view of the BESS-Polar II experiment.}
\label{fig9}
\end{center}
\end{figure}
Figure~\ref{fig9} shows the general layout of the BESS-Polar II spectrometer~\cite{yos09}. Based on feedback from analysis of the previous flight, various improvements were done and many of the detectors and systems were re-designed and re-fabricated both to improve performance and flight duration. The BESS apparatus is built around a superconducting magnet surrounded by detectors needed for particle identification. 

In 2013 the BESS collaboration presented the measurement of energy spectra of primary protons in the range 0.2-120 GeV by BESS-Polar II. In this energy range, the modulation of cosmic rays in the heliosphere is a relevant and fundamental problem in space plasma physics, heliospheric physics and in cosmic ray physics. This is a very active field, which complements state-of-the-art numerical models (e.g.~\cite{pot13}) with new data collected by long duration experiments. Since the heliosphere is the only astrophysical system in which in-situ spacecraft measurements are available, its study can also lead to fundamental insights applicable to larger astrophysical systems. 

\begin{figure}[!hbt]
\begin{center}
\includegraphics[width=7.5cm]{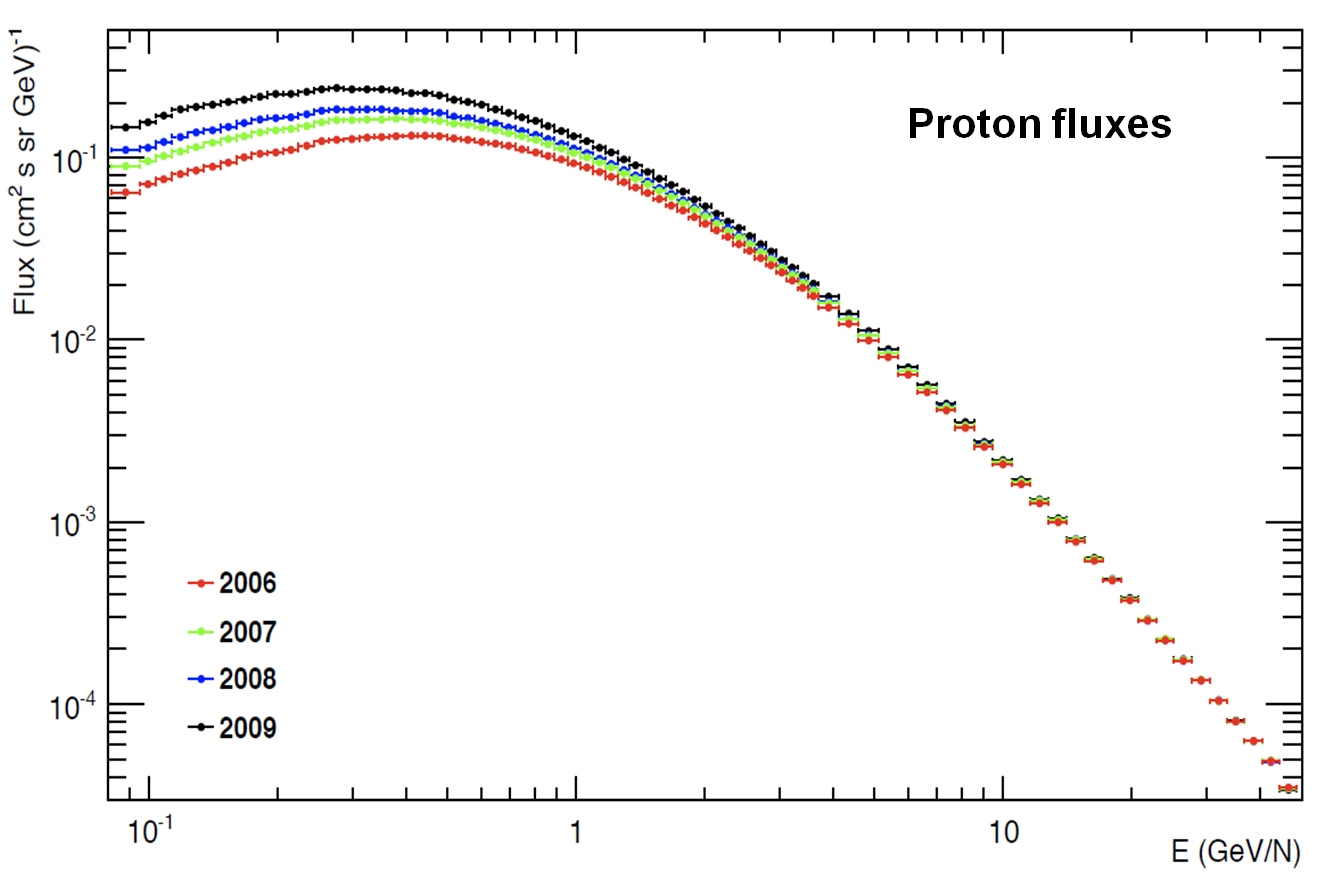}
\caption{Proton energy spectrum as function of the time measured by PAMELA.}
\label{fig10}   
\end{center}
\end{figure}

The solar wind significantly affects the low energy part of cosmic rays, as it can be easily verified monitoring the time variation of the cosmic-ray fluxes and their dependence with the sun activity. 
Figure~\ref{fig10}
 shows low energy proton flux measured by PAMELA from July 2006 till December 2009~\cite{adr13}. The flux variation with time can be clearly seen as well as the decreasing significance of the variation as the energy increases, becoming negligible above 5~GeV. BESS-Polar II results concerning the proton flux variation as function of time have been presented at the ICRC 2013~\cite{sak13}. From the figures presented by the BESS collaboration (e.g. figure 6 in \cite{sak13}), it can be seen PAMELA and BESS contemporary results show an excellent agreement.

\section{Cosmic rays composition}
\label{chemic}
Several long duration balloon flights have been carried out in the last decade to study the chemical composition of cosmic rays at energies higher than hundreds of GeV.

The Cosmic Ray Energetics And Mass (CREAM) experiment was flown six times using long duration balloon-borne. 
\begin{figure}[!hbt]
\begin{center}
\includegraphics[width=3.2cm]{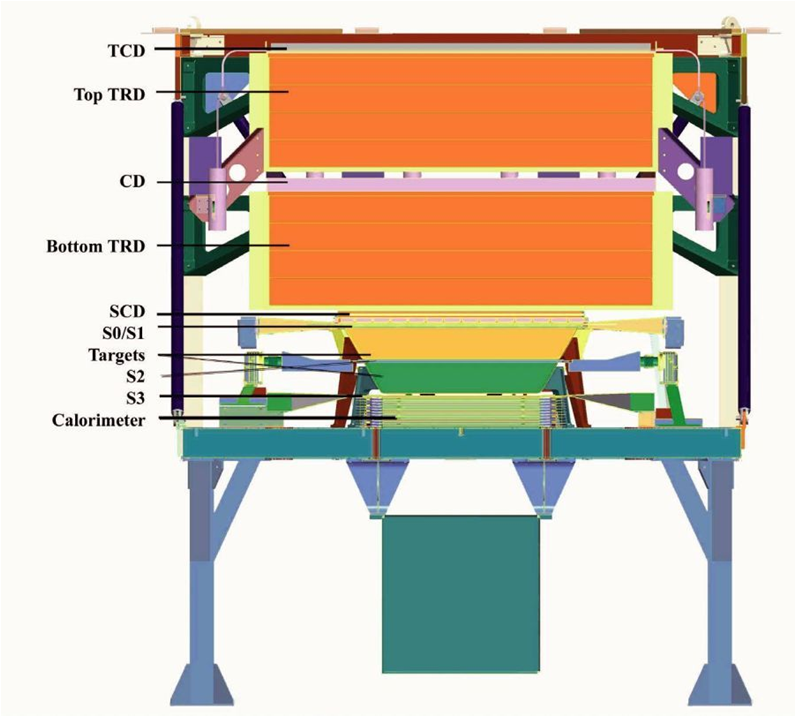}
\includegraphics[width=3.2cm]{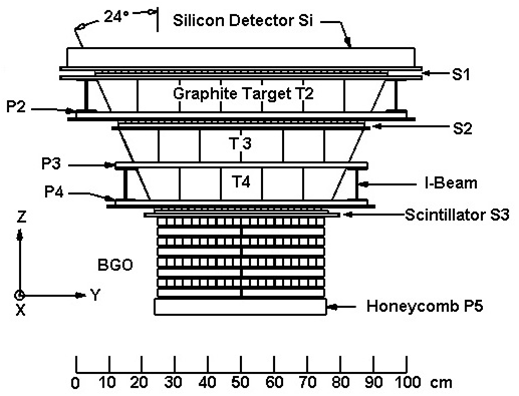}
\caption{On the left, a schematic view of the CREAM-II apparatus. On the right, a schematic view of ATIC-2.}
\label{fig1112}   
\end{center}
\end{figure}
Figure~\ref{fig1112} on the left shows the CREAM-II configuration. Carbon targets are used to start the protons and nuclei showers. The particle energy is then measured by using a calorimeter module. Timing charge detectors, Cherenkov detectors and TRDs are used to determine the velocity and to measure the electrical charge of the particles. Another apparatus that was flown from Antarctica three times is the Advanced Thin Ionization Calorimeter (ATIC), figure~\ref{fig1112} on the right~\cite{wef08}. As the name suggests, also in this detector the energy is measured with a deep calorimeter after the forced interaction of nuclear component of cosmic rays in graphite targets. One of the main results of these two experiments is the measurement of high energy proton and helium nuclei fluxes. 
\begin{figure}[!hbt]
\begin{center}
\includegraphics[width=7.5cm]{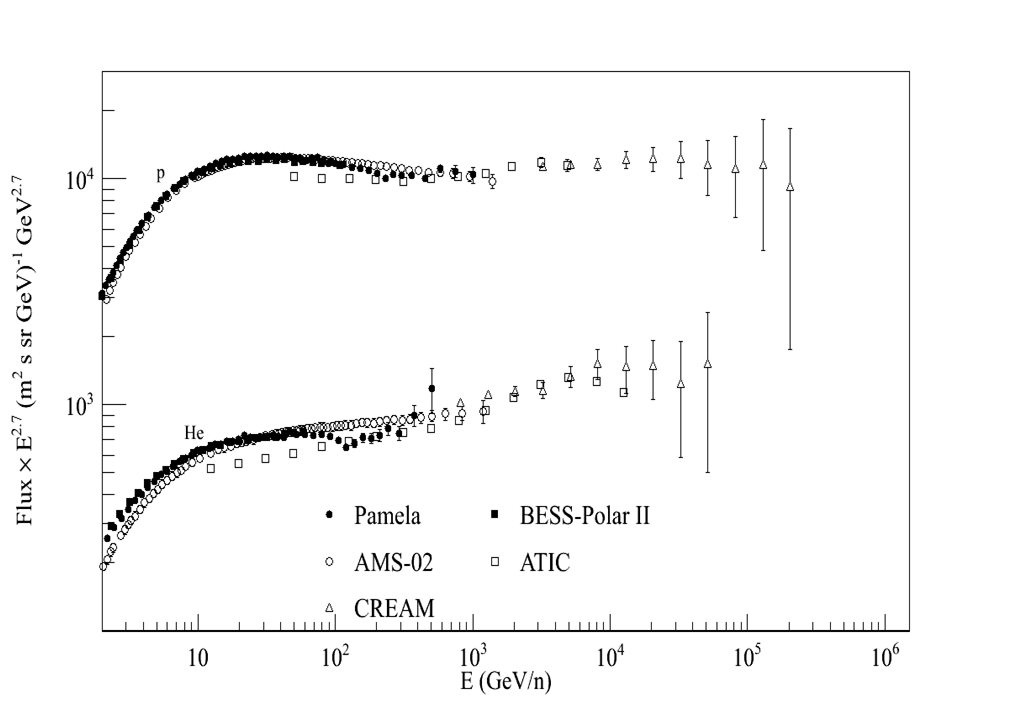}
\caption{Proton and helium nuclei spectra as measured by PAMELA, AMS-02 (preliminary), BESS-Polar II (preliminary), CREAM and ATIC-2.}
\label{fig13}   
\end{center}
\end{figure}
Figure~\ref{fig13} shows these two fluxes as measured by PAMELA~\cite{adr11a}, AMS-02 (preliminary data)~\cite{ams13a}, BESS-Polar II (preliminary data)~\cite{sak13}, CREAM~\cite{yoo11}, and ATIC-2~\cite{pan09}. Considering systematic and statistical uncertainties, the overall agreement is good, both from the point of view of the absolute normalization and from the point of view of the spectral shape. As discussed previously, the present generation of experiment is probably not precise enough to draw conclusions about thin spectral features, however there seems to be a general agreement about two characteristics of these fluxes. First of all, there is strong and convincing evidence that the spectral index of proton differs from the helium nuclei ones. Secondly, there is an indication of a hardening of proton and helium nuclei spectra not only in the PAMELA data themselves but also by combining measurements obtained by calorimeters and spectrometers. 

There is an indication of a similar spectral index change also in the measurement of nuclei with charge greater than helium as measured by CREAM~\cite{ahn09} and the Transition Radiation Array for Cosmic Energetic Radiation (TRACER). TRACER is a balloon borne detector for the measurements of single element heavy cosmic ray nuclei (boron to iron) in the energy range from 10$^{13}$ to several 10$^{15}$ eV per nucleus~\cite{ave09}. A schematic view of the TRACER detector 
\begin{figure}[!hbt]
\begin{center}
\includegraphics[width=7.5cm]{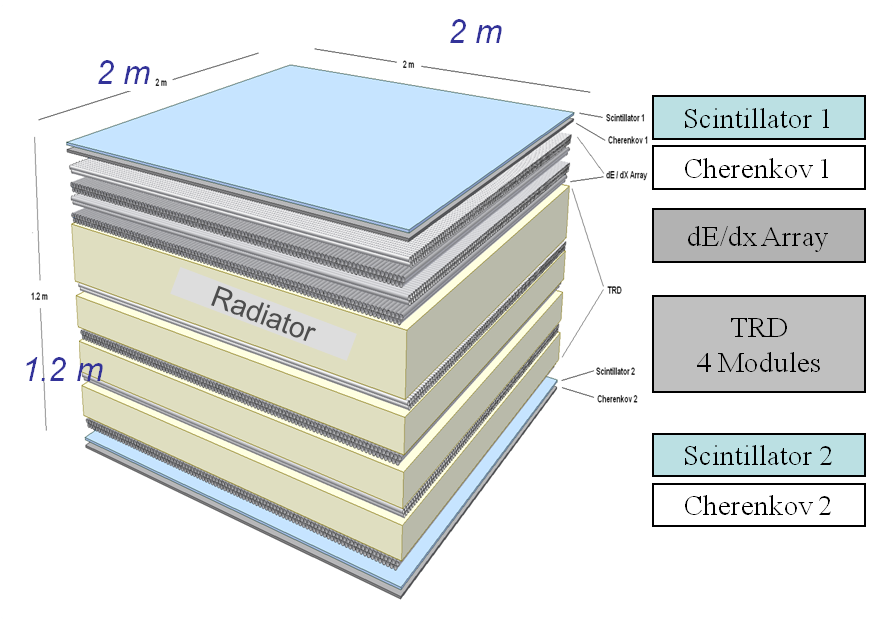}
\includegraphics[width=7.5cm]{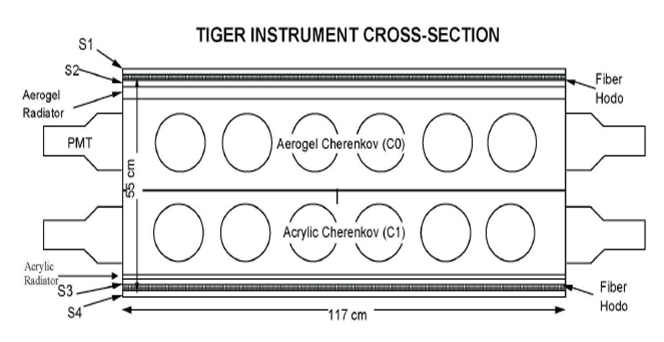}
\caption{Top panel, the TRACER detector. Bottom panel, the TIGER experiment~\cite{rau09}.}
\label{fig1415}   
\end{center}
\end{figure}
is shown in figure~\ref{fig1415}, upper panel. The main components are two Cherenkov detectors, four TRD modules and a ionization loss array detector. Nuclei energy measurement is performed by making use of these three different detectors in the three particle velocity windows in which the signal response is proportional to the particle incoming energy. Excellent charge discrimination is obtained by selecting particles as function of their velocity, their ionization losses and their signal on the two different Cherenkov detectors. TRACER was flown three times, from Ft. Sumner in the USA, from Antarctica and from Sweden. Results are shown in figure~\ref{fig16} 
\begin{figure}[!hbt]
\begin{center}
\includegraphics[width=7.5cm]{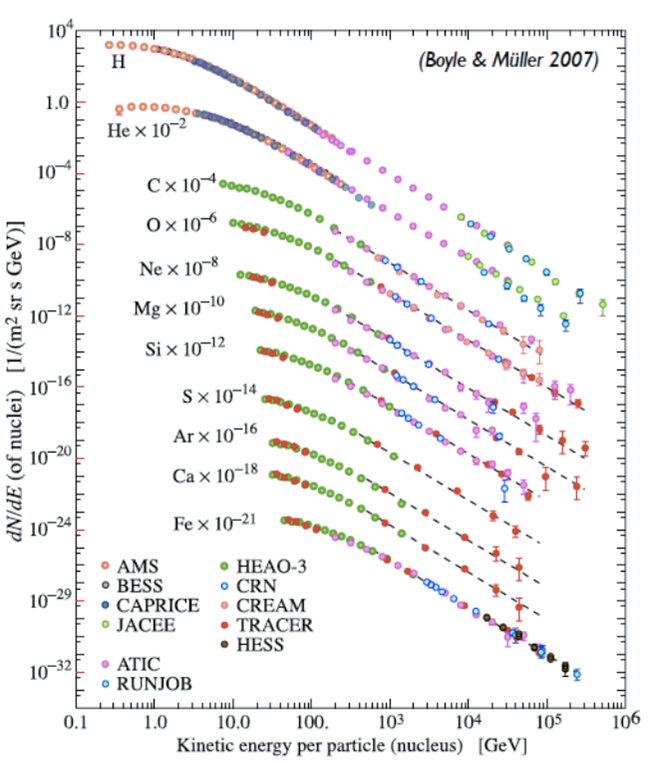}
\caption{Flux as a function of energy for the major components of the primary cosmic radiation. The dashed line represents a power-law fit to the TRACER data~\cite{ave09} above 20~GeV. TRACER results are compared to other experimental data sets \cite{boe03,hai04,yoo11,pan09,mul91,alc00a,apa99,asa98,eng90}.}
\label{fig16}   
\end{center}
\end{figure}
compared to data from other experiments~\nocite{boe03,hai04,yoo11,pan09,mul91,alc00a,apa99,asa98,eng90}. The TRACER measurements are in agreement with recent results from CREAM and ATIC and extend towards the knee without reaching it. The results indicate that power-law fits in energy to the measured spectra can be made with the same power-law exponent ($\gamma=2.67\pm0.05$), for all elemental species~\cite{obe11}. Similarly, the CREAM group reported a common power-law exponent of $\gamma=2.66\pm0.04$.

Nuclei with even higher charge are measured by the Trans-Iron Galactic Element Recorder (TIGER)~\cite{rau09} and its updated version SUPERTIGER~\cite{bin14}. The TIGER experiment is designed to measure cosmic rays with atomic number greater than 26 (Iron). TIGER was launched twice from Antarctica, while SUPERTIGER was launched in December 2012 and completed a 55 days balloon flight, thereby setting a record for the longest successful scientific balloon flight. The TIGER detector is shown in figure~\ref{fig1415}, bottom panel. Two Cherenkov detectors with different radiators are the core of the instrument. The charge of incoming particles is measured by combining ionization loss and Cherenkov measurements. The TIGER results allow the study of the composition of the sources of cosmic rays, indicating that a mixed combination of standard Solar System material and Massive Star ejecta is needed to fit the data. Moreover there is an indication that dust grains undergo a more effective acceleration with respect to interstellar matter gas, supporting a model in which the galactic cosmic ray sources are OB associations~\cite{rau09}.

\section{Electrons flux}
Electrons constitute only about 1\% of the total cosmic ray flux. This component provides important information regarding the origin and propagation of cosmic rays in the Galaxy. In fact, because of their low mass, electrons undergo severe energy losses through synchrotron radiation in the Galactic magnetic field and inverse Compton scattering with the ambient photons. More than in the case of the cosmic ray nuclear component, structures in the spectral shape of the electron energy spectrum are expected as a contribution of large energy losses and, possibly, of the new sources.

In the recent years new experimental results from ground based, balloon-borne and from satellite-based experiments have been presented. Especially the results from the space-borne Fermi, PAMELA and AMS experiments have been particularly significant. Fermi is a gamma ray space telescope~\cite{ack10}, made by a precise tracking system, an anticoincidence and an electromagnetic calorimeter. A dedicated charged particle trigger combined with the use of the calorimeter allows Fermi to measure the leptonic component of the cosmic rays with good precision.
\begin{figure}[!hbt]
\begin{center}
\includegraphics[width=7.5cm]{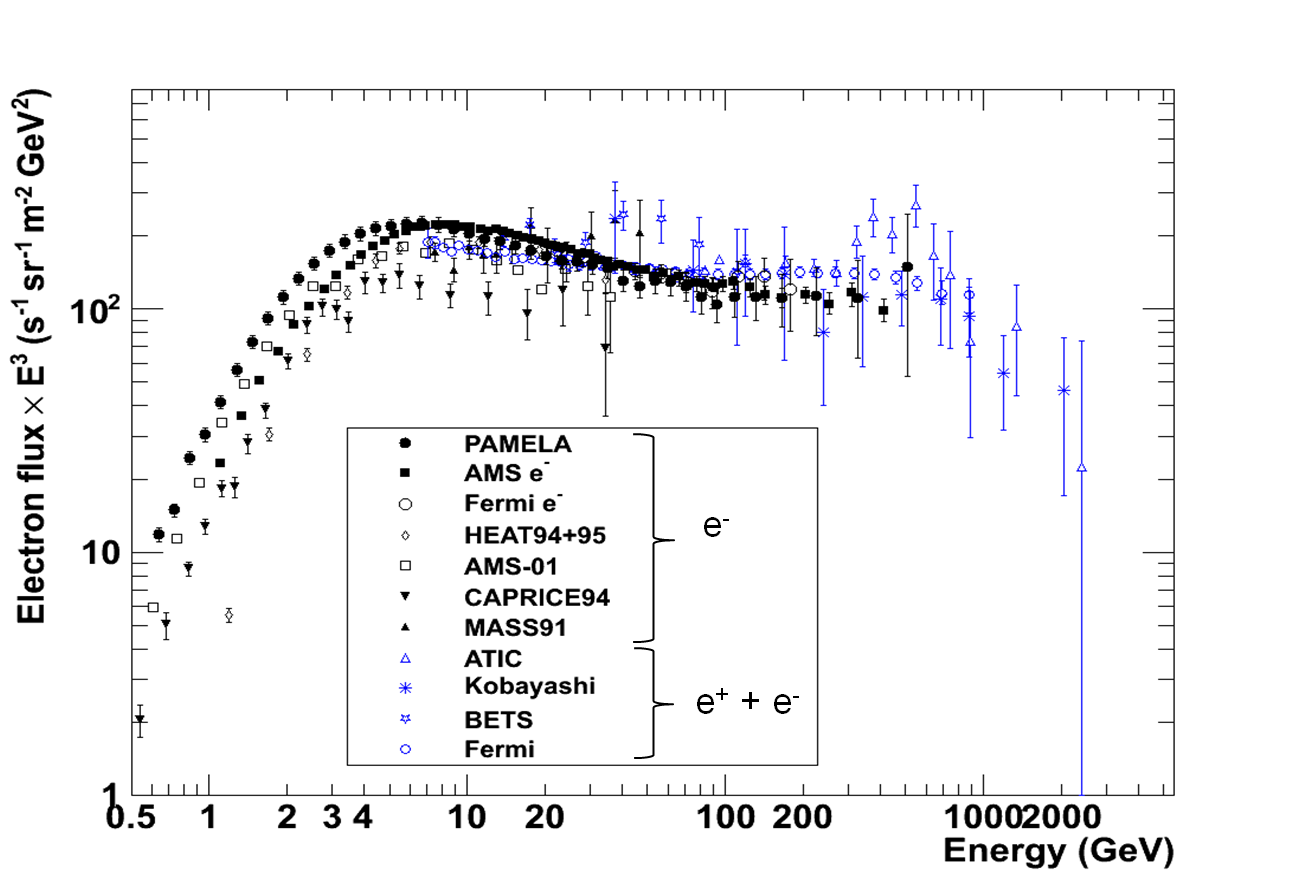}
\caption{Electron spectrum. Blue symbols represent data sets in which electrons and positrons are jointly measured~\cite{cha08,kob99,ack10,tor01}, black symbols represent the negative electron flux~\cite{adr11b,boe00,duv01,alc00b,gri02,ams13c,ack12}.}
\label{fig17}   
\end{center}
\end{figure}
Results from several experiments are shown in figure~\ref{fig17}. Black symbols represent the pure negative electron spectrum as measured by previous generation experiments~\cite{boe00,duv01,alc00b,gri02} and from the most recent PAMELA~\cite{adr11b}, Fermi~\cite{ack12} and AMS-02~\cite{ams13c}. Blue symbols show the calorimetric measurement of the sum of electrons and positrons~\cite{cha08,kob99,ack10,tor01}. It can be noticed that the peak observed by ATIC at hundreds of GeV is not observed by the other experiments. However there is a good agreement between the most recent measurements and a certain flattening of the spectrum can be observed above some tens of GeV. At energies around 10 GeV it can be noticed how the systematic uncertainties play a significant role when they are dominant with respect to statistical ones. In fact, the all electron spectrum as measured by Fermi~\cite{ack10} is lower, below about 30 GeV, with respect to the negative electron spectrum measured by the same experiment~\cite{ack12} while the two spectra are in agreement at higher energies. The two results are consistent when considering systematic uncertainties (not shown in figure), however it is evident that the systematic uncertainty estimated in the Fermi all electron spectrum has a distortion effect on the resulting flux.

\section{Antiparticles}
\begin{figure}[!hbt]
\begin{center}
\includegraphics[width=7.5cm]{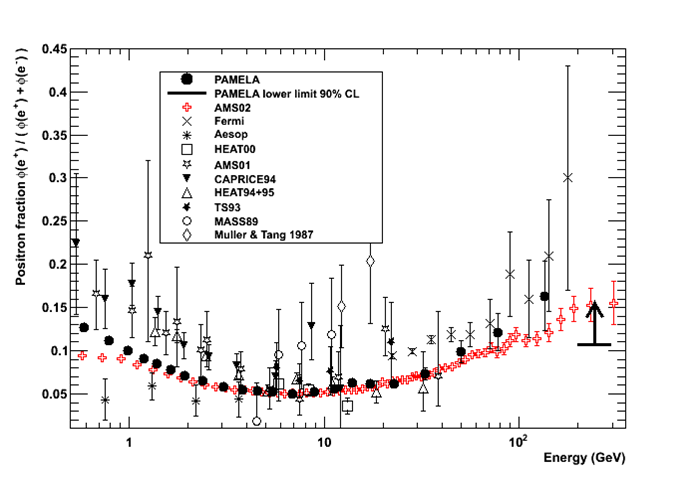}
\caption{Positron fraction measured by PAMELA~\cite{adr13b}, Fermi~\cite{ack12} and AMS~\cite{agu13}, compared to previous measurements~\cite{boe00,alc00b,gol96,gas06,bar97,cle07,mul87,bea93}.}
\label{fig18}   
\end{center}
\end{figure}

\begin{figure}[!hbt]
\begin{center}
\includegraphics[width=7.5cm]{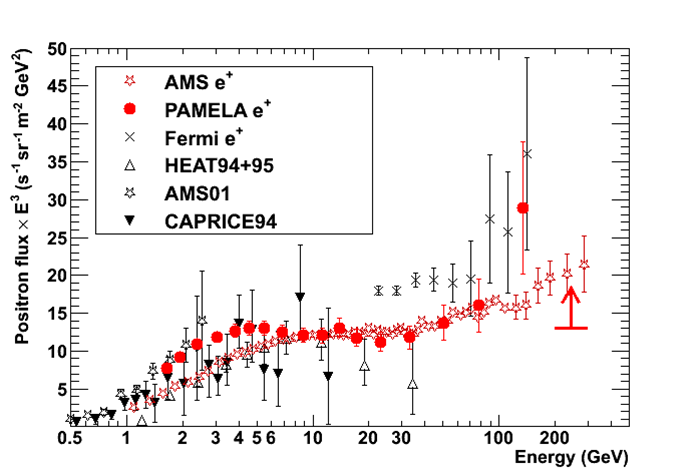}
\caption{Positron flux spectrum~\cite{boe00,duv01,alc00b,ams13c,ack12,adr13b}.}
\label{fig19}   
\end{center}
\end{figure}

Antiparticle measurement is the main goal of the present day space experiments dedicated to the cosmic ray studies. 
Figure~\ref{fig18} shows 
a comparison of the latest positron fraction measurements~\cite{agu13,ack12,adr13b} with previous measurements~\cite{boe00,alc00b,gol96,gas06,bar97,cle07,mul87,bea93}. It can be noticed how the latest generation experiment is able to provide a much higher statistics, covering a large energy range. The agreement between PAMELA and AMS-02 is excellent, considering the effects of solar modulation, while the Fermi results, obtained by using the Earth magnetic field East-West effect to determine the sign of the charge, are higher but with large systematic uncertainties (not shown in figure). With respect to the proton and helium fluxes, in this case the statistics is large but at high energy the statistical error is comparable to the estimated systematic as can be seen also in the positron flux spectrum, figure~\ref{fig19}.
In this case, there is a very nice agreement between PAMELA and AMS-02 data not only in the fraction of fluxes but also in the measurement of the pure particle spectrum. AMS-02 and Fermi results confirm that the increase noticed in the positron fraction is due to a harder positron spectrum at high energy. Future measurement of the positron fluxes at higher energy made by AMS-02 could give a contribution in understanding the origin of high energy positrons.

\begin{figure}[!hbt]
\begin{center}
\includegraphics[width=7.5cm]{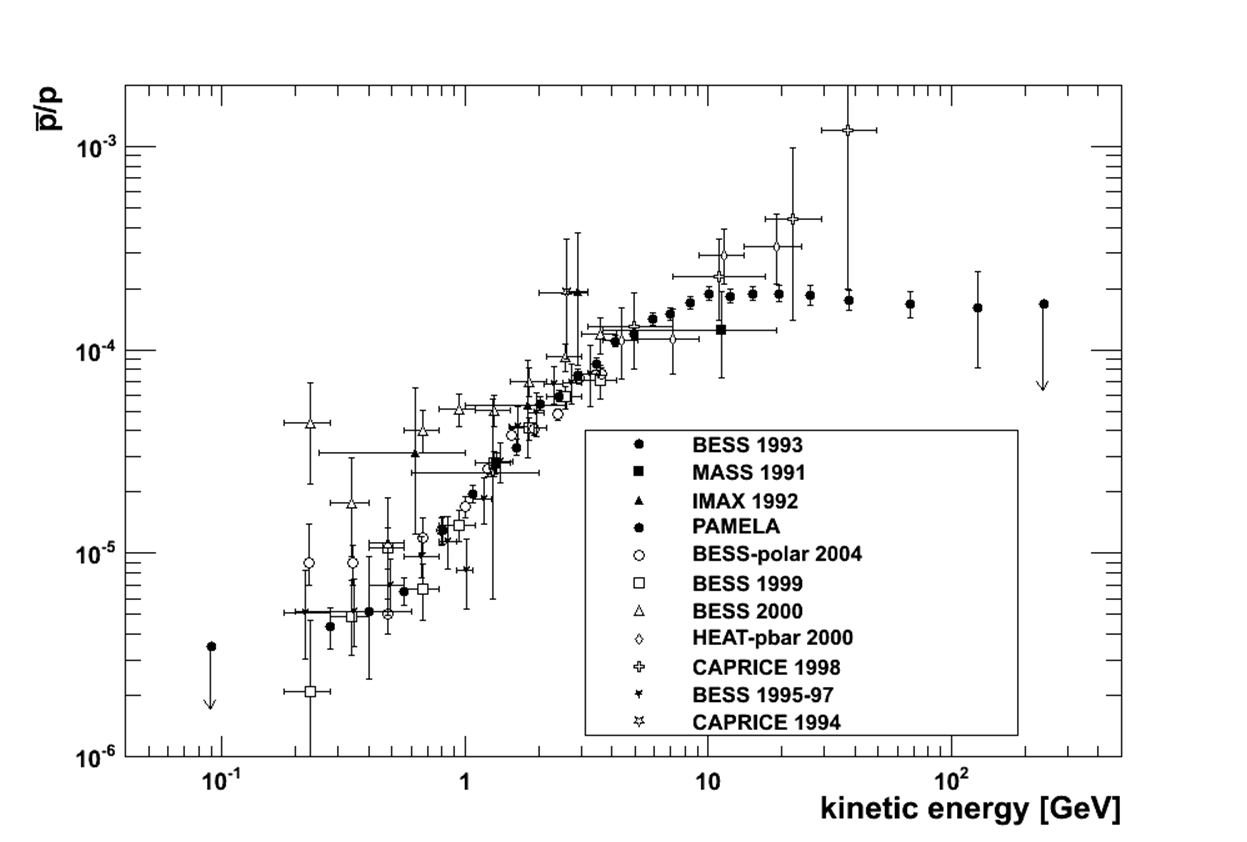}
\caption{Antiproton to proton ratio measured by PAMELA~\cite{adr12}, BESS~\cite{asa02,abe08,ued96,mat96} and older experiments~\cite{bea01,boe97,boe01,hof96,mit96}.}
\label{fig20}   
\end{center}
\end{figure}

AMS-02 results will also include the antiproton flux and antiproton to proton ratio measurements. Figure~\ref{fig20} shows 
the present status of the antiprotons to proton ratio. Latest PAMELA results~\cite{adr12} are compared to BESS measurements at low energy~\cite{asa02,abe08,ued96,mat96} where BESS provide the best statistics and to older experiments~\cite{bea01,boe97,boe01,hof96,mit96}. The antiproton measurements are in agreement with a pure secondary production of these particles via the interaction of primary cosmic rays with the interstellar matter. This measurement combined with the positron one, that instead shows an unexplained increase, represents a challenging puzzle for cosmic ray models. A source of leptons must be implemented in the model or very peculiar and ``leptophylic'' models of dark matter are required to explain the increase in the positron measurements together with the ``standard'' model results of the antiproton flux.

\section{Future experiments}
Future experiments aim to continue the study of cosmic rays by detecting electrons, nuclei and anti-particles both at high and low energy.

CALET is an experiment designed to measured the all electron cosmic ray spectrum from 20 GeV to 10 TeV\~cite{tor07}. The apparatus is built around a very deep electromagnetic calorimeter of 30 radiation length. It will be placed on board the International Space Station (ISS) in fiscal year 2014. CALET will significantly extend previous electron measurement.

A different approach has been proposed for the CREST experiment that aims to detect the synchrotron photons generated at x-ray energies by TeV cosmic-ray
electrons in the Earth magnetic field~\cite{sch07}. While affected by a relatively poor energy resolution, the experiment will be sensitive to electrons of energies greater than 2 TeV, extending the electron measurements to the multi-TeV region. The apparatus, carried on a long duration balloon-flight, will
be able to observe up to 30 electrons with energy greater than 2 TeV in a 2 week flight.

A calorimetric approach will be employed by the Gamma-400 experiment~\cite{gal13}. This will be a dual experiment aimed to study both the high-energy gamma-ray flux and the charged cosmic rays, both electrons and light nuclei. The apparatus will be placed on board a Russian satellite, which launch is foreseen for 2018. With a deep and large calorimeter (acceptance of about 1 m$^2$sr), Gamma-400 should be able to extend significantly the cosmic ray measurements performed by CALET and to measure the nuclear component of cosmic rays toward the knee.

An extension of the CREAM long duration balloon program is the ISS-CREAM mission~\cite{seo13}, to be delivered to the International Space Station (ISS) and externally mounted at the Japanese Experiment Modules Exposed Facility (JEM-EF) KIBO. ISS-CREAM presents many challenges to the development team: the 1.200 kg estimated mass of the payload is over twice the mass of any previously launched payload using the JAXA’s HTV. The team will modify the existing instruments to meet the new requirements of the launch vehicle and ISS.

The detection of low energy antideuterons produced in WIMP–WIMP annihilation is the main goal of the general antiparticle spectrometer (GAPS) experiment~\cite{hai09}. GAPS has substantial discovery potential for dark matter within the minimal supersymmetric model and its extensions, and models with universal extra dimensions. GAPS is designed to be a balloon-borne experiment.

\section{Conclusions}
The latest generation of cosmic ray particle detectors has brought and is bringing many exciting results. Proton, helium and, possibly, highly charged nuclei spectra seems to harden at similar rigidities. Moreover, there is a strong indication that proton and helium nuclei have indeed a different spectral index. 

The measurement of antiparticles in the cosmic rays has been very popular in the last years, with a possible indication of dark matter detection in the positron fraction has to take into account not only the ``missing signal'' in the antiproton measurement but also the possibility of nearby astrophysical sources capable of accelerating electrons and positrons.

All these measurements are challenging the cosmic ray standard model. New results from current and future experiments will probably contribute in developing a more precise description of the sources, acceleration and propagation of cosmic rays. 

A special care, however, must be used when interpreting experimental data: thanks to larger acceptances and acquisition time, it is likely that systematic uncertainties will dominate in a big part of the detected energy window. In such cases, it is always important to carefully describe the sources of these uncertainties and their effects on the measurements. Unlikely statistical errors, systematic uncertainties are estimated and they are strongly related to the experimental apparatus and their effect can bring not only renormalization problems but also distortions in the flux measurements.

\bibliographystyle{elsarticle-num}
\bibliography{mocchiutti_crbtsm}

\end{document}